\documentclass[twocolumn,prl,aps,psfrag]{revtex4} 
\usepackage{amsmath,amssymb,epsf,bm}
\usepackage{dcolumn}
\usepackage{graphicx} 
\usepackage[percent]{overpic}
\usepackage[usenames,dvipsnames]{color}
\usepackage{psfrag,latexsym} 
\usepackage{epstopdf}
\usepackage[caption=false]{subfig}
\usepackage{comment}
\usepackage{ulem}

\begin{document}

\title{Spin superfluidity and long-range transport in thin-film ferromagnets}

\author{Hans Skarsv{\aa}g}
\email{hans.skarsvag@ntnu.no}
\affiliation{Department of Physics, Norwegian University of Science and
Technology, NO-7491 Trondheim, Norway}

\author{Cecilia Holmqvist}
\affiliation{Department of Physics, Norwegian University of Science and
Technology, NO-7491 Trondheim, Norway}

\author{Arne Brataas}
\affiliation{Department of Physics, Norwegian University of Science and
Technology, NO-7491 Trondheim, Norway}

\date{\today}

\begin{abstract}
In ferromagnets, magnons may condense into a single quantum state. Analogous to superconductors, this quantum state may support transport without dissipation. Recent works suggest that longitudinal spin transport through a thin-film ferromagnet is an example of spin superfluidity. Although intriguing, this tantalizing picture ignores long-range dipole interactions; we demonstrate that such interactions dramatically affect spin transport.
In single-film ferromagnets, "spin superfluidity" only exists at length scales (a few hundred nanometers
in yttrium iron garnet) somewhat larger than the exchange length.
Over longer distances, dipolar interactions destroy spin superfluidity.
Nevertheless, we predict re-emergence of spin superfluidity in tri-layer ferromagnet--normal metal--ferromagnet films of $\sim 1\, \mu$m in size. Such systems also exhibit other types of long-range spin transport in samples several micrometers in size.
\end{abstract}

\maketitle
% % % % % % % % % % % % % % % % % % % % % % % % % % % % % % % % % % % % % % % % % % % % % % % % % % % % % % % % % % % % % % % % % % % % % % % % % % % % % % % % % % % % % % % % 

When matter enters a superfluid phase, it behaves like a fluid with zero viscosity and can support currents without dissipation.
It has been suggested that certain ferromagnets may exhibit spin superfluidity (SSF) \cite{sonin:advphys10,konig:prl02,nogueira:epl04}. The superfluid spin-drag properties induced by spin transfer and spin pumping (SP) in a normal metal--ferromagnet--normal metal system have recently been computed \cite{takei:prl14,chen:prb14,takei:arxiv2015}. Related studies have also explored Josephson spin currents between magnons condensates \cite{nakata:prb14}. Experimental studies have suggested that the temporal decrease of magnon condensates is associated with SSF \cite{Clausen:arXiv1503.00482}. 

In the absence of magnetic fields, SSF is indeed an intriguing possibility because its realization would allow spin currents to propagate without significant losses over long distances. These spin transport properties may be useful for low-dissipation interconnects, spin logic devices, and non-volatile magnetic memory devices. Our work demonstrates that SSF can exist in thin-film ferromagnetic systems, but two ferromagnets (rather than one) are required to cancel long-range dipole interactions. We do not observe signatures of long-range SSF in single-film ferromagnets.

Recent works have hypothesized that easy-plane ferromagnetic thin films exhibit SSF. In such systems, a monotonously precessing magnetization leads to meta-stable spin-current-carrying states whose topological properties protect against dissipation \cite{sonin:advphys10}.
Spin relaxation induces a finite resistance proportional to the system size \cite{takei:prl14}. Nevertheless, ferromagnetic insulators (FIs) have exceptionally low spin dissipation rates, and the spin supercurrent decays over a large length scale. Furthermore, the spin-relaxation-induced algebraic decay of the spin supercurrent significantly differs from the exponential decay of the spin current carried by spin waves \cite{hillebrands:apl14}. Although magnetic anisotropy destroys the linear SSF response, the spin current is predicted to flow with negligible dissipation when the bias is sufficiently large \cite{takei:prl14,chen:prb14}.

It is well known that long-range dipole interactions dramatically affect the spin-wave dispersion in thin films \cite{kalinikos:jphys86,serga:jpd10}. Low-energy magnons strongly interact, and the coupling between them decreases algebraically as they spatially separate. Magnon interactions also influence Bose-Einstein condensation such that the condensate occurs at a finite wavevector around the magnon energy minimum \cite{demokritov:nature06}. Naturally, the long-range nature of dipole interactions can also strongly affect the SSF. However, for SSF to be useful, it must exist over long, hopefully macroscopic, length scales.

In the previous theoretical investigations of SSF presented in Refs.~\onlinecite{sonin:advphys10,konig:prl02,nogueira:epl04,takei:prl14,chen:prb14,takei:arxiv2015}, the approximate dipole field was included as an easy-plane anisotropy. However, the dipole interaction also has a dynamical component not included in Refs.~\onlinecite{sonin:advphys10,konig:prl02,nogueira:epl04,takei:prl14,chen:prb14,takei:arxiv2015}. It is the long-range nature of this component that qualitatively changes the dispersion of magnons \cite{kalinikos:jphys86}. When the system is smaller than the exchange length, the energy associated with the exchange stiffness dominates, and the system may exhibit SSF. However, the exchange length in FIs, such as yttrium iron garnet (YIG), is $\sim 20$ nm, and dipole interactions become increasingly important at larger length scales.

\begin{figure}
\vspace{-1cm}
\hspace{-4mm}
\begin{overpic}
[width=7cm]{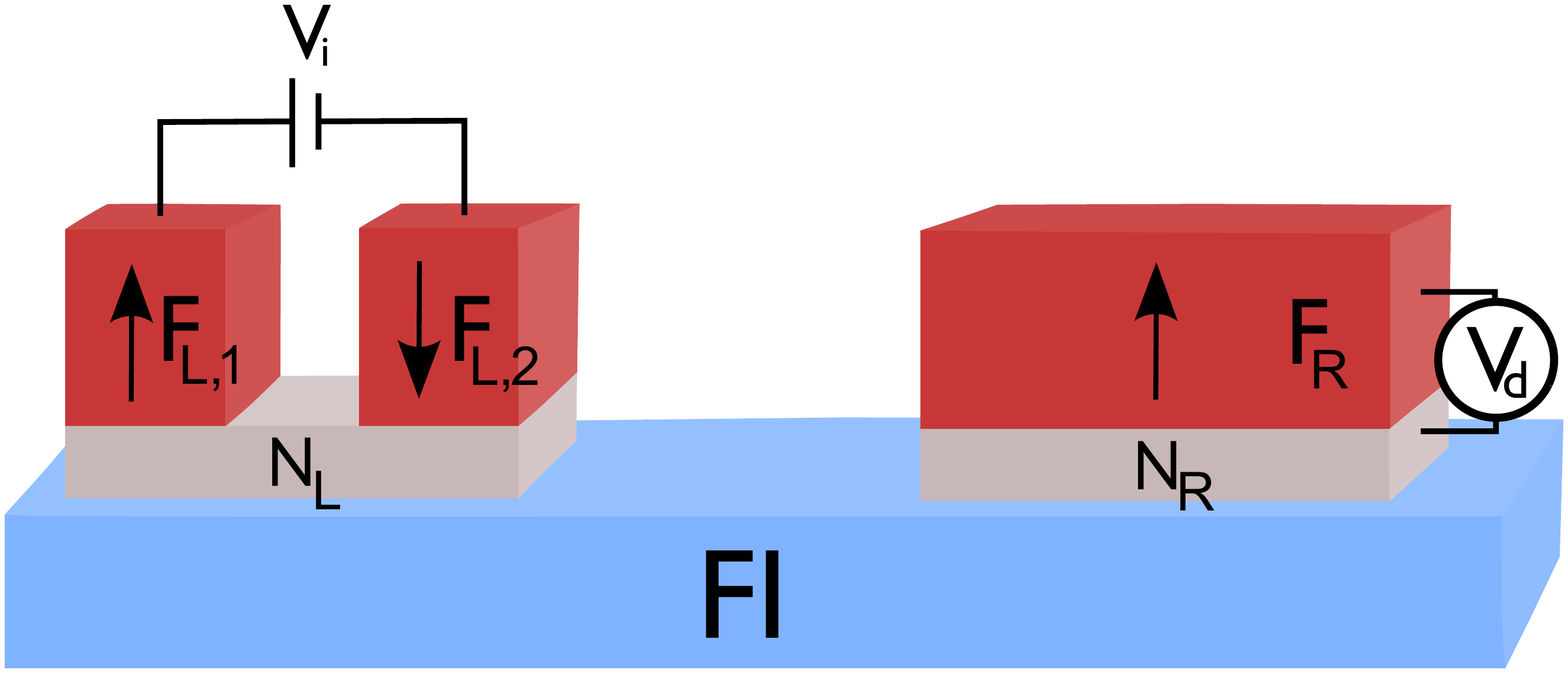}
\put(0,18){(a)}
\end{overpic}
\vspace{-1.1cm}\\
\begin{overpic}
[width=8cm]{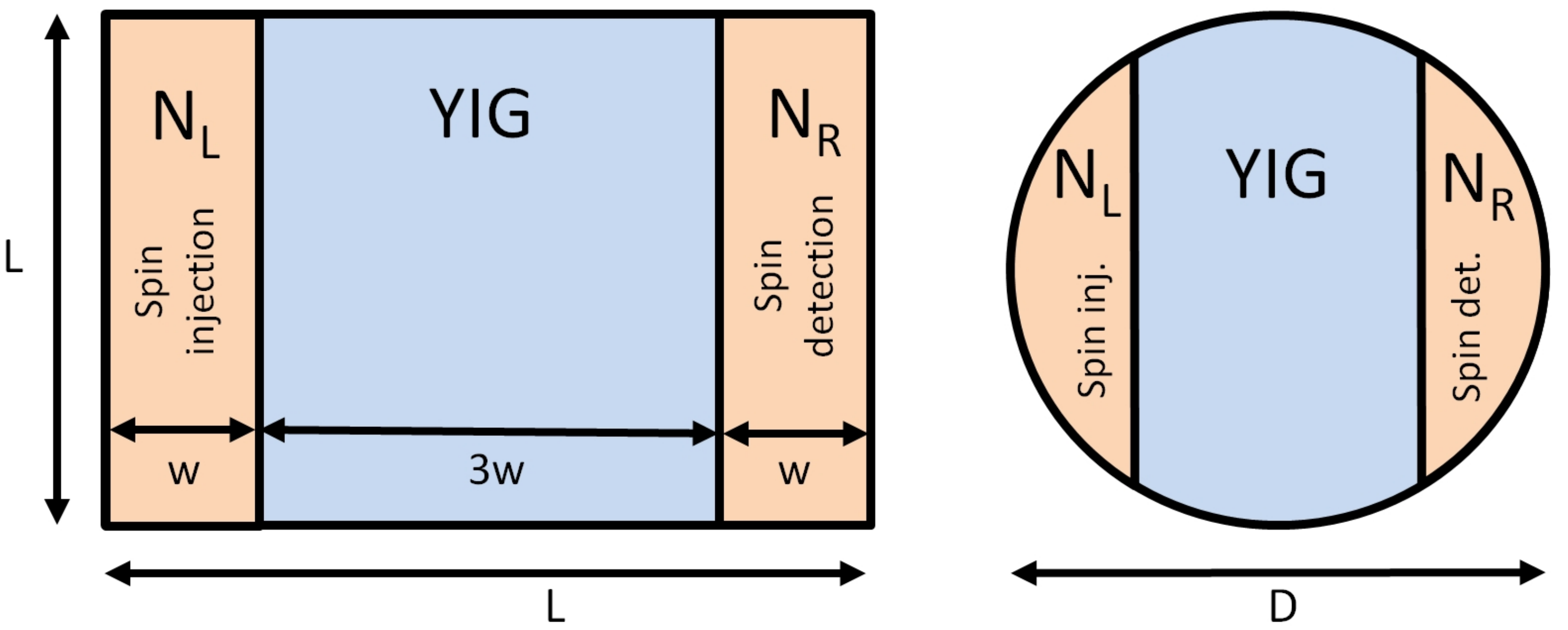}
\put(0,0){(b)}
\put(60,0){(c)}
\end{overpic}
\caption{(color online). (a) A square FI thin film in contact with a spin injector (top left) and a spin detector (top right). A bias voltage $V_i$ injects a spin current via a lateral spin valve in which the ferromagnetic leads have a perpendicular magnetic anisotropy.
SP from the FI thin film into the detector induces a voltage $V_d$. (b) The width of the contacts, $w$, is $20\%$ of the total length of the FI sample, $L$. (c) Circular disk with diameter $D$. The spin injection and detection contacts each cover $20\%$ of the disk area.
\label{fig:system}}
\end{figure}

In this Letter, we investigate the complete effect of dipole interactions on spin transport through an FI thin film. We consider both square and circular devices, as shown in Fig.~\ref{fig:system}. As expected, dipole interactions completely alter the spin transport properties. We find that "SSF" can only be achieved when the system size is on the order of the exchange length, which implies that SSF is not a useful method for transporting spin information across sizeable distances. For example, in YIG, which is a widely used FI because of its low dissipation, SSF occurs only across distances of a few hundred nanometers.
In comparison, typical spin-wave propagation lengths may reach $\sim 5~\mathrm{mm}$ in YIG logic devices \cite{chumak:natcomm14}.
Moreover, because of dipole-induced anisotropies, a sufficiently high spin accumulation bias is required to induce a spin current. As with the spin resistance across the sample,
this spin-accumulation threshold strongly depends on the geometry of the system and increases with the system size.

Nonetheless, the concept of SSF in ferromagnetic systems remains useful, but not in single films, as previously envisaged. Instead of the single-film configuration, we propose a tri-layer structure. Exchange coupling between two FIs via a normal metal can secure an anti-parallel configuration of the magnetizations in the two ferromagnets. We demonstrate that such synthetic antiferromagnets maintain long-range SSF over distances much greater than the exchange length. We also show that even when the two films differ, a spin supercurrent and ultimately a long-range non-SSF spin current can flow over sufficiently long distances in typical realizations.

The setup in Ref.~\onlinecite{takei:prl14} nicely illustrates SSF behavior. The
spin Hall effect leads to spin injection. In turn, spin-transfer torque (STT) causes the magnetization to precess, thereby leading to SP out of the opposite contact. This SP is detected via the inverse spin Hall effect. This geometry therefore requires the contacts to be attached to the thin sides of the FI.  The resulting resistance per area can be expressed in the form of an Ohm's law using the interface resistances and an internal resistance,
\begin{equation}
r=1/g_L^{\perp} + 1/g_R^{\perp} + r_\alpha,
\label{eq:Ohm}
\end{equation}
where $g^{\perp}_L$ and $g^{\perp}_R$ are the transverse ("mixing") interface spin conductances, and the internal spin-relaxation-induced resistance is $r_\alpha=g_\alpha/(g_R^{\perp} g_L^{\perp})$, where $g_\alpha=2e^2M_sL\alpha_0/\hbar^2\gamma$. Here, $M_s$ is the saturation magnetization, 
$\gamma$ is the gyromagnetic ratio, $\alpha_0$ is the intrinsic Gilbert damping coefficient, and $L$ is the system length. The system exhibits SSF because the internal resistance $r_\alpha$ vanishes when the spin is conserved ($\alpha_0 \rightarrow 0$). With spin dissipation, the internal resistance increases algebraically with the length of the system.
 
To further utilize SSF, we suggest using a larger injection area with a spin valve attached to the top of the FI; see Fig.~\ref{fig:system}(a). Ignoring dipole interactions,
Ohm's law (\ref{eq:Ohm}) remains valid, but the intrinsic conductance becomes
\begin{equation}
g_\alpha=\frac{4\pi M_s\mathcal{V}\alpha_0}{\hbar\gamma \mathcal{A}_\text{c}}\frac{e^2}{h}, 
\label{eq:gintrinsic}
\end{equation}
where $\mathcal{V}$ is the volume of the FI and $\mathcal{A}_\text{c}$ is the injection/detection contact area. From Eqs.\ \eqref{eq:Ohm} and \eqref{eq:gintrinsic}, one can conclude that the SSF can be made arbitrarily long-range by increasing the injection area in proportion to the length between the detector and injector contacts. Without dipole interactions, the SSF is limited only by the contact conductances, and a spin current can flow over
macroscopic lengths. However, as discussed below, dipole interactions dramatically reduce the applicability of this finding.

In the geometry employed herein, a
spin current is injected using the left contact (L), which consists of a spin valve with
two ferromagnets, F$_{\text{L},1/2}$, that exhibit perpendicular magnetic anisotropy and are coupled to a normal metal, N$_\text{L}$; see Fig.~\ref{fig:system}(a). We can calculate the injected spin accumulation in $\mathrm{N_\text{L}}$, $\boldsymbol{\mu}_L=\mu_L\hat{z}$, using circuit theory \cite{brataas:prl00}. Assuming an effective conductance across the N$_\text{L}$$|$FI interface, $\tilde{g}_\text{FI}$, and low spin memory loss in N$_\text{L}$, we find that $\mu_L=eV_i(g_\uparrow+g_\downarrow)/(g_\uparrow+g_\downarrow+\tilde{g}_\text{FI})$. Here, $g_{\uparrow(\downarrow)}$ is the conductance of the majority (minority) electrons across the two F$_{\text{L},1/2}|$N$_\text{L}$ interfaces.

This spin accumulation then drives the FI dynamics of the local magnetization direction, $\mathbf{m}(\mathbf{r},t)$, at position $\mathbf{r}$ and time $t$.
The spin angular momentum transported through the FI thin film is subsequently detected by the right contact (R), which consists of a normal metal, $\mathrm{N_R}$, connected to a ferromagnet, F$_\text{R}$. The spin accumulation pumped into $\mathrm{N_R}$ is given by $\boldsymbol{\mu}_R(\boldsymbol{r})=-\hbar \mathbf{m}\times \dot{\mathbf{m}}|_{\boldsymbol{r}\in \text{R}}$ and can be measured according to the voltage, $V_d$, across the $\mathrm{N_R}$$|$F$_\text{R}$ junction.

At low temperatures, SSF can be described semi-classically \cite{sonin:advphys10,takei:prl14,chen:prb14,takei:arxiv2015}. The magnetization dynamics are then described by the Landau-Lifshitz-Gilbert (LLG) equation,
\begin{equation}\label{eq:llg}
\dot{\mathbf{m}}=-\gamma \mathbf{m}\times \mathbf{H}_\text{eff}+\alpha \mathbf{m}\times \dot{\mathbf{m}}-\alpha' \mathbf{m}\times\mathbf{m}\times \boldsymbol{\mu}/\hbar,
\end{equation}
where, in the left (right) contact region,
the spin accumulation $\boldsymbol{\mu}=\boldsymbol{\mu}_{L(R)}$ and $\alpha'=\alpha_{L(R)}$;
both quantities are zero otherwise.
The dimensionless parameter $\alpha_{L(R)}=g^{\perp}_{L(R)} \hbar^2\gamma/2e^2M_sd$, where $d$ is the FI  thickness. The local Gilbert damping coefficient is $\alpha=\alpha_0+\alpha'$, where $\alpha'$ is the spin pumping enhancement. The effective field, $\mathbf{H}_\text{eff}$, consists of the exchange field, $\mathbf{H}_\text{ex}=2A/(M_s) \nabla^2 \mathbf{m}$, where $A$ is the exchange constant, and the dipole field, $\mathbf{H}^\text{tot}_\text{dip}$, which fulfills Maxwell's equations in the magnetostatic approximation,
\begin{equation}
\nabla\times \mathbf{H}^\text{tot}_\text{dip}=0, \hspace{5mm} \nabla \cdot (\mathbf{H}^\text{tot}_\text{dip}+4\pi M_s\mathbf{m})=0.
\end{equation}
The dipole field is related to the local magnetization by Green's functions: $\mathbf{H}^\text{tot}_\text{dip}=\int_\mathcal{V} \mathrm{d}^3 \mathbf{r}' \widehat{G}(\mathbf{r}-\mathbf{r'})~\mathbf{m}(\mathbf{r'},t)$, where $\widehat{G}$ is a 2nd-rank tensor whose elements are $G_{\alpha \beta}=-(1/4 \pi) \partial^2_{\alpha \beta'} (1/| \mathbf{r}-\mathbf{r}' | )$ \cite{gurevich:96}.
We consider an FI thinner than the magnetic exchange length, $d\lesssim l_\text{ex}=\sqrt{A/2\pi M_s^2}$, 
such that any variation of $\mathbf{m}$ across the thickness is negligible.
Then, one can divide the total dipole field into an easy plane term, $\mathbf{H}_\text{EP}=-4 \pi M_s m_z \hat{z}$, and the remainder of the dipole field, $\mathbf{H}_\text{dip}$.

A consequence of the dipole field $\mathbf{H}_\text{dip}$ is that
the spin-wave eigenspectrum strongly depends on the spin-wave propagation direction relative to the magnetization \cite{serga:jpd10}.
At long wavelengths and no applied magnetic field, spin waves propagating with wave vector $k_\parallel$ parallel to the magnetization are exchange dominated,
and their frequency is $\omega(k_\parallel)=\gamma k_\parallel \sqrt{8\pi A}$. 
In the perpendicular configuration, the spin waves are governed by the  dipole interaction and $\omega(k_\perp)=4\pi M_s \gamma \sqrt{2k_\perp d}$, where $k_\perp$ is the perpendicular wave vector.
SSF is associated with steady-state solutions of Eq.~(3) where the magnetization has a small out-of-plane component and performs 2$\pi$ precessions \cite{sonin:advphys10}. Hence, the relative orientation of the transport direction and the magnetization alternates between the exchange-dominated and dipole-dominated regimes.
Non-local dipole interactions are therefore of a crucial importance.
However, the full inclusion of these interactions transforms the LLG equation \eqref{eq:llg} into
a complicated 2$^\text{nd}$-order non-linear integro-differential equation in time and the in-plane coordinates,
and finding its solution requires considerable numerical efforts.
For this purpose, we performed graphics processing unit (GPU)-accelerated micromagnetic simulations on several computers over a long time period \cite{vansteenkiste:14}.

We first consider a square YIG thin film. The square geometry results in two dipole-induced easy axes that extend diagonally across the sample in addition to the easy plane anisotropy.
The injection and detection contacts cover $20 \%$ of the thin film's surface area, as shown in Fig.~\ref{fig:system}(b).
Hence, the ratio $\mathcal{A}_\text{c}/\mathcal{V}=1/(5d)$ that controls the internal conductance \eqref{eq:gintrinsic} is independent of the length $L$. We further neglect any spin-memory loss inside the contacts because of the long spin-diffusion length of the Cu contacts, $l^{\text{Cu}}_\text{sf}=100-1500~\mathrm{nm}$ \cite{villamor:prb2013}.
Regarding the Cu$|$YIG interface, spin-pumping experiments have measured transverse (mixing) conductance values in the range of $g_\text{Cu$|$YIG}^{\perp}h/e^2\sim 10^{13-15}~\mathrm{cm}^{-2}$ \cite{wang:prl14,du:prappl14}. We choose $g^{\perp}_{L/R}h/e^2= 5\cdot10^{14}~\mathrm{cm}^{-2}$, which, combined with $d_\text{YIG}=5~\mathrm{nm}$, yields $\alpha_{L/R}\approx 0.01$.
We also use $4\pi M_s=1750$ G \cite{serga:jpd10}, $A=3.7\cdot 10^{-7}$ erg/cm \cite{klingler:jpd15}, and $\alpha_0=1\cdot 10^{-3}$ \cite{wang:prl14,du:prappl14}.
The $\mathrm{N_R}|$F$_\text{R}$ interface is assumed to be a tunnel interface; therefore, the SP current across the FI$|\mathrm{N_R}$ is compensated by the STT generated by $\boldsymbol{\mu}_R$.
Generally, $\boldsymbol{\mu}_R$ contains both ac and dc components. We denote the z component of the dc spin accumulation in the right contact averaged over the contact area by $\langle \mu_R^z \rangle$ and investigate its behavior as a function of $\mu_L$ and the system size.

\begin{figure}[t]
\begin{overpic}
[width=0.99\columnwidth,clip=true]{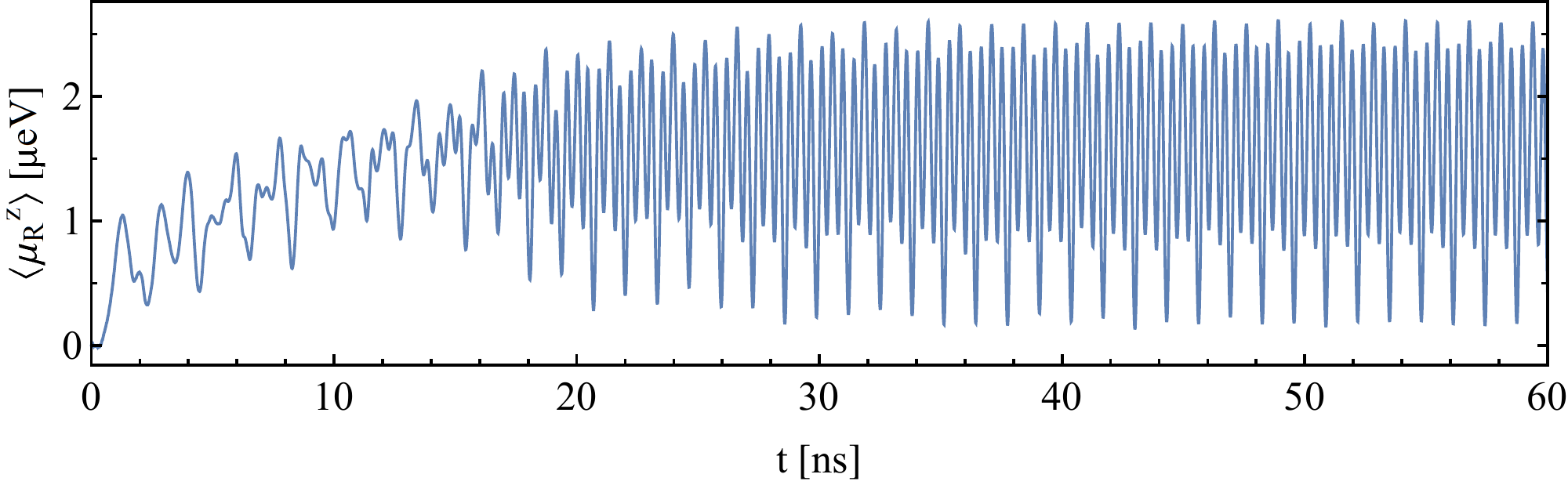}
\put(0,3){(a)}
\end{overpic}
\begin{overpic}
[width=0.49\columnwidth,clip=true]{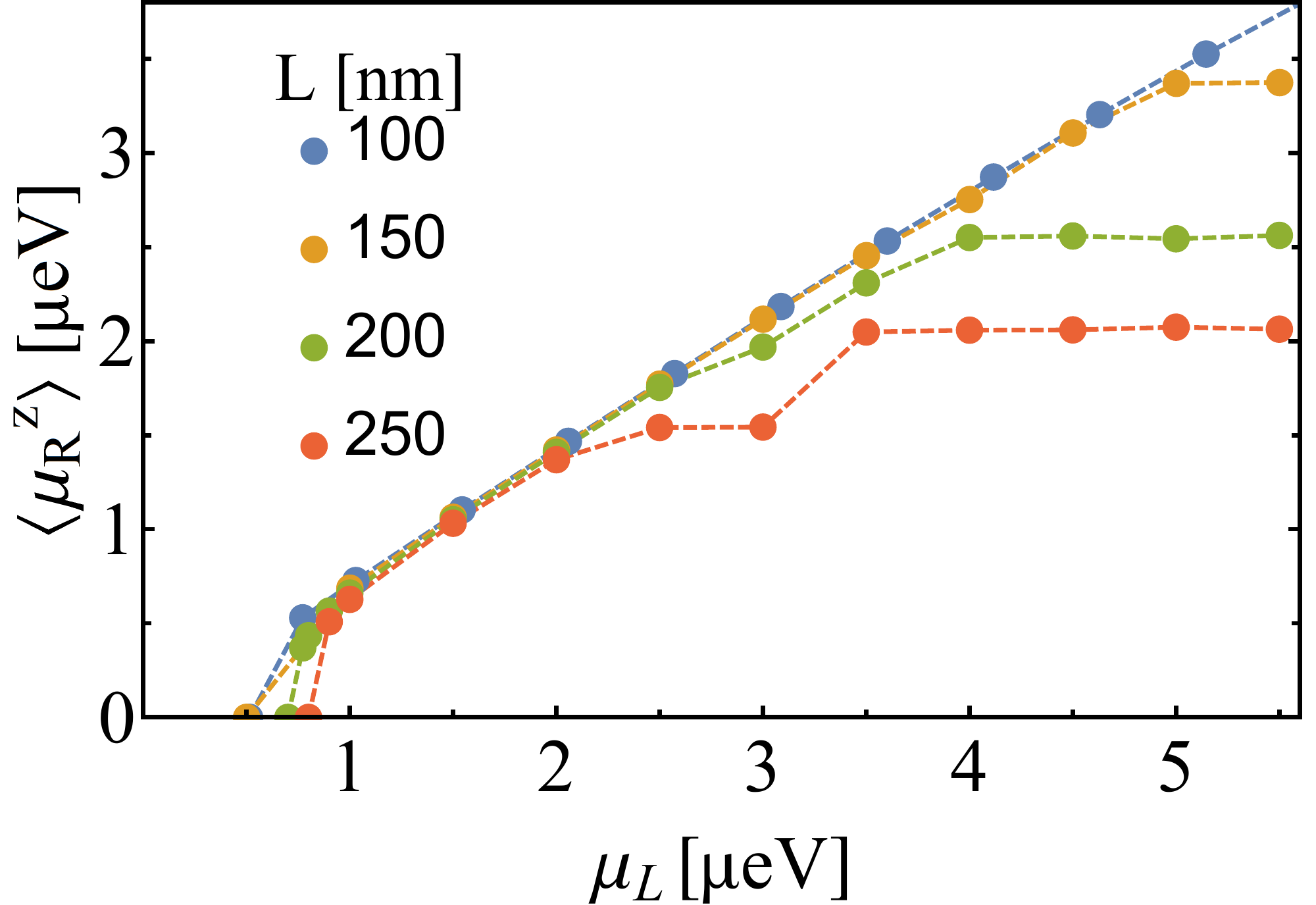}
\put(0,3){(b)}
\end{overpic}
\begin{overpic}
[width=0.49\columnwidth,clip=true]{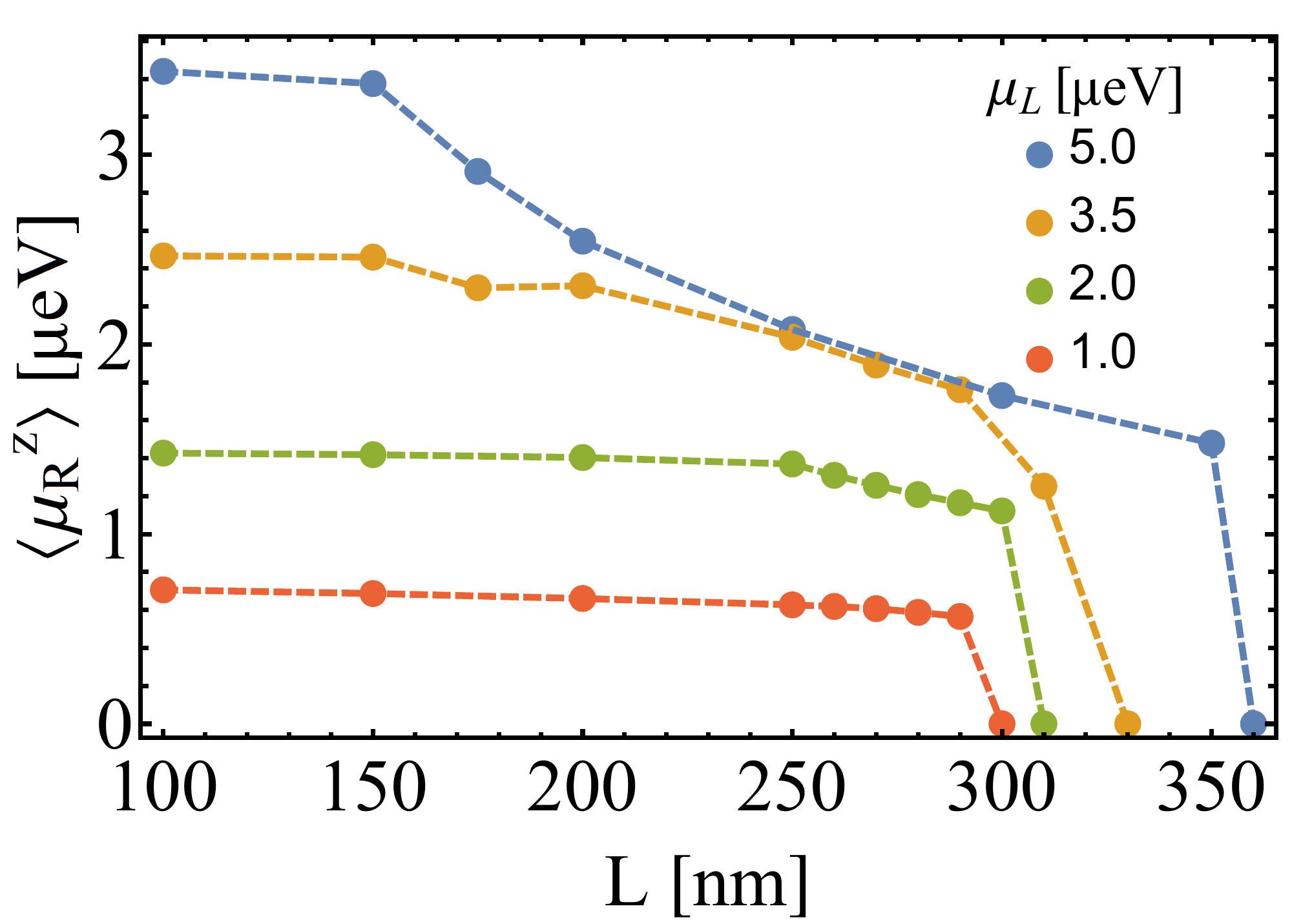}
\put(0,3){(c)}
\end{overpic}
\caption{(color online). (a) Time evolution of $\mu_R^z$ when $L=250$ nm and $\mu_L=3.0$ $\mu$eV.
(b) The average dc spin accumulation in the right contact, $\langle \mu^z_R \rangle$, as a function of the average spin accumulation in the left contact, $\mu_L$, in samples with fixed lengths, $L$, and square geometries. (c) $\langle \mu^z_R \rangle$ as a function of $L$.\label{fig:plots}}
\end{figure}

In the micromagnetic simulations, we start in a uniform state and let this state evolve into a steady state; see Fig.~\ref{fig:plots}(a).
Figure~\ref{fig:plots}(b) shows the dc spin accumulation in the steady state, $\langle \mu^z_R\rangle$, as a function of $\mu_L$. For small values of $\mu_L$, $\langle \mu^z_R\rangle$ is finite only above a threshold value, $\mu_L^{\text{th}}$, which increases with increasing system size, $L$. Above the threshold, $\langle \mu^z_R\rangle$ increases linearly with $\mu_L$ until it saturates at a maximum value, $\mu_L^{\text{sat}}$.
The vanishing $\langle \mu^z_R\rangle$ for $\mu_L <\mu_L^{\text{th}}$ is due to the dipole field that, for sample sizes $\gtrsim \l_{\text{ex}}$, creates a shape anisotropy sufficiently large to pin the magnetic moments of the sample.
For $\mu_L \gtrsim \mu_L^{\text{th}}$, $\langle \mu^z_R \rangle$ is close to the ideal value given by Eqs.~\eqref{eq:Ohm} and \eqref{eq:gintrinsic}, thus indicating the emergence of SSF.
The saturation of $\langle \mu^z_R \rangle$ at large values of $\mu_L$ is caused by the interplay of the shape anisotropy and non-local magnon-magnon interactions.
This saturation is analogous to the leveling off of the cone angle in a ferromagnetic resonance experiment as a function of applied power due to magnon-magnon-interaction-mediated Suhl instabilities \cite{suhl:pcs57,bahlmann:jap96}.
Note that because $\mu_L^{\text{sat}}$ decreases with increasing $L$, the interval in which SSF is possible, $\mu_L^{\text{th}}< \mu_L<\mu_L^{\text{sat}}$, shrinks with increasing $L$ until $\mu_L^{\text{sat}} < \mu_L^{\text{th}}$. This "squeezing" effect restricts SSF to samples less than a few hundred nanometers in size.

In Fig. \ref{fig:plots}(c), $\langle \mu^z_R\rangle$ is plotted as a function of system length, $L$. For small system sizes, $\langle \mu^z_R\rangle$ is independent of $L$, as in the SSF Eqs.~\eqref{eq:Ohm} and \eqref{eq:gintrinsic}.
For intermediate film sizes, the dynamic dipole field dominates the exchange interaction. Although spin transport is still possible, it is not mediated by the SSF.
For large sample sizes, dipole pinning suppresses any spin transport.
This result demonstrates that the SSF is restricted to samples smaller than 300 nm in size and that coherent longitudinal spin transport over macroscopic distances in a single-layer sample with a square geometry is impossible.

\begin{figure}[t]
\begin{overpic}
[width=0.47\columnwidth,clip=true]{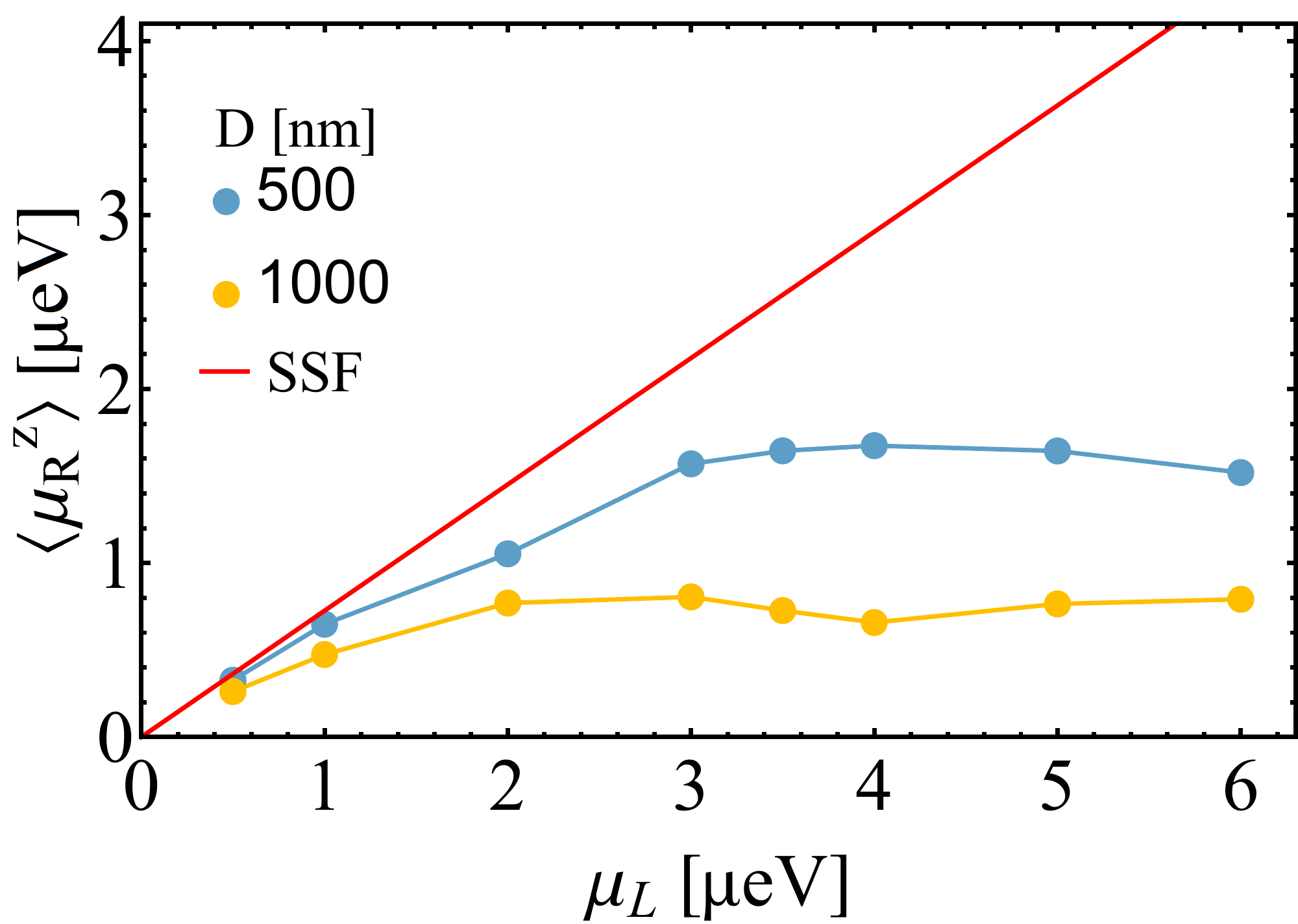}
\put(0,3){(a)}
\put(38,59){\includegraphics[width=1cm]{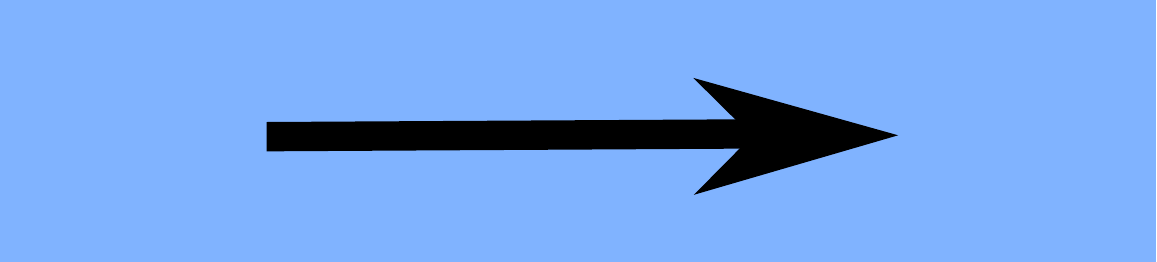}}
\end{overpic}
\begin{overpic}
[width=0.49\columnwidth,clip=true]{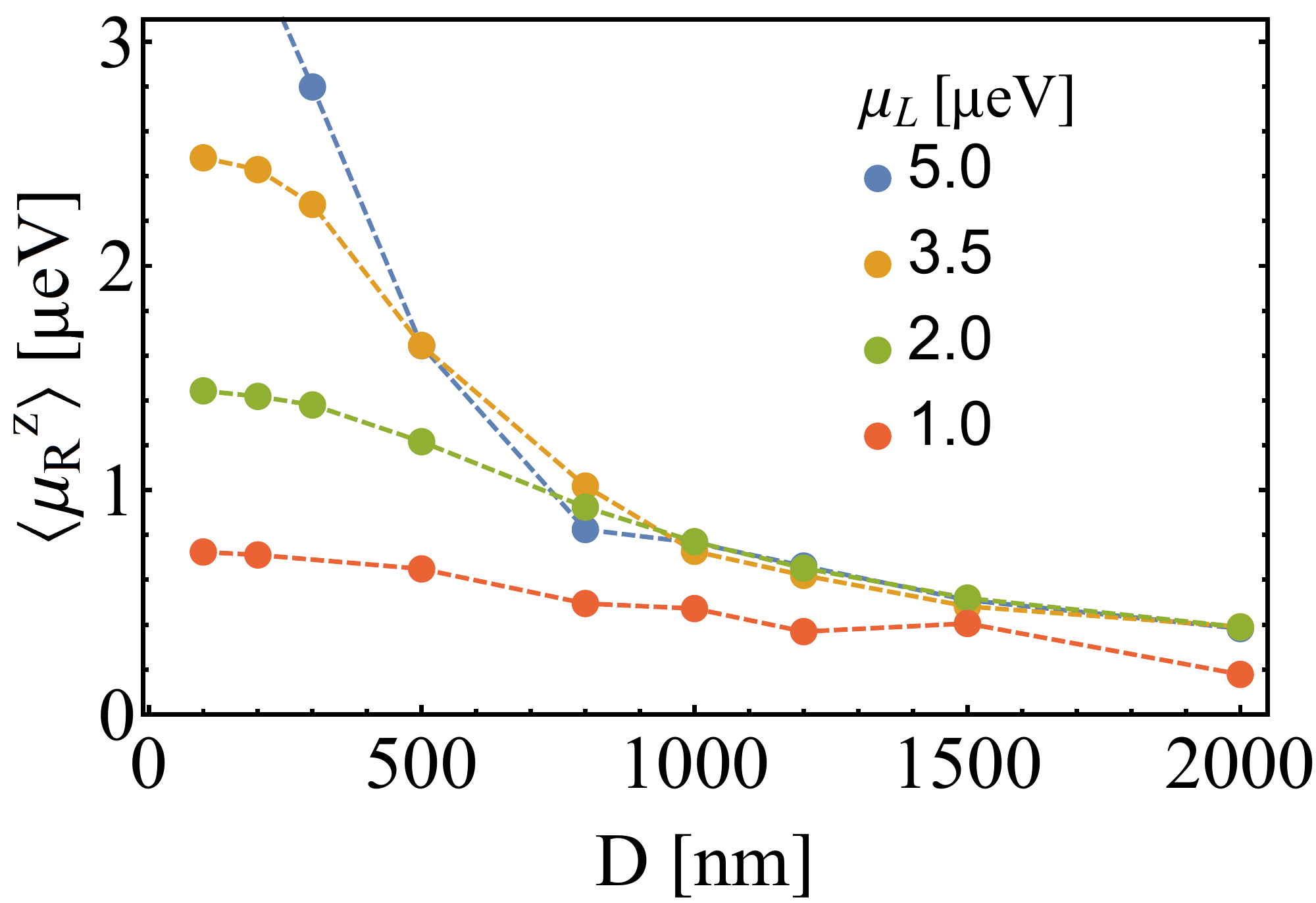}
\put(0,3){(b)}
\end{overpic}
\vspace{0.2cm}
\begin{overpic}
[width=0.47\columnwidth,clip=true]{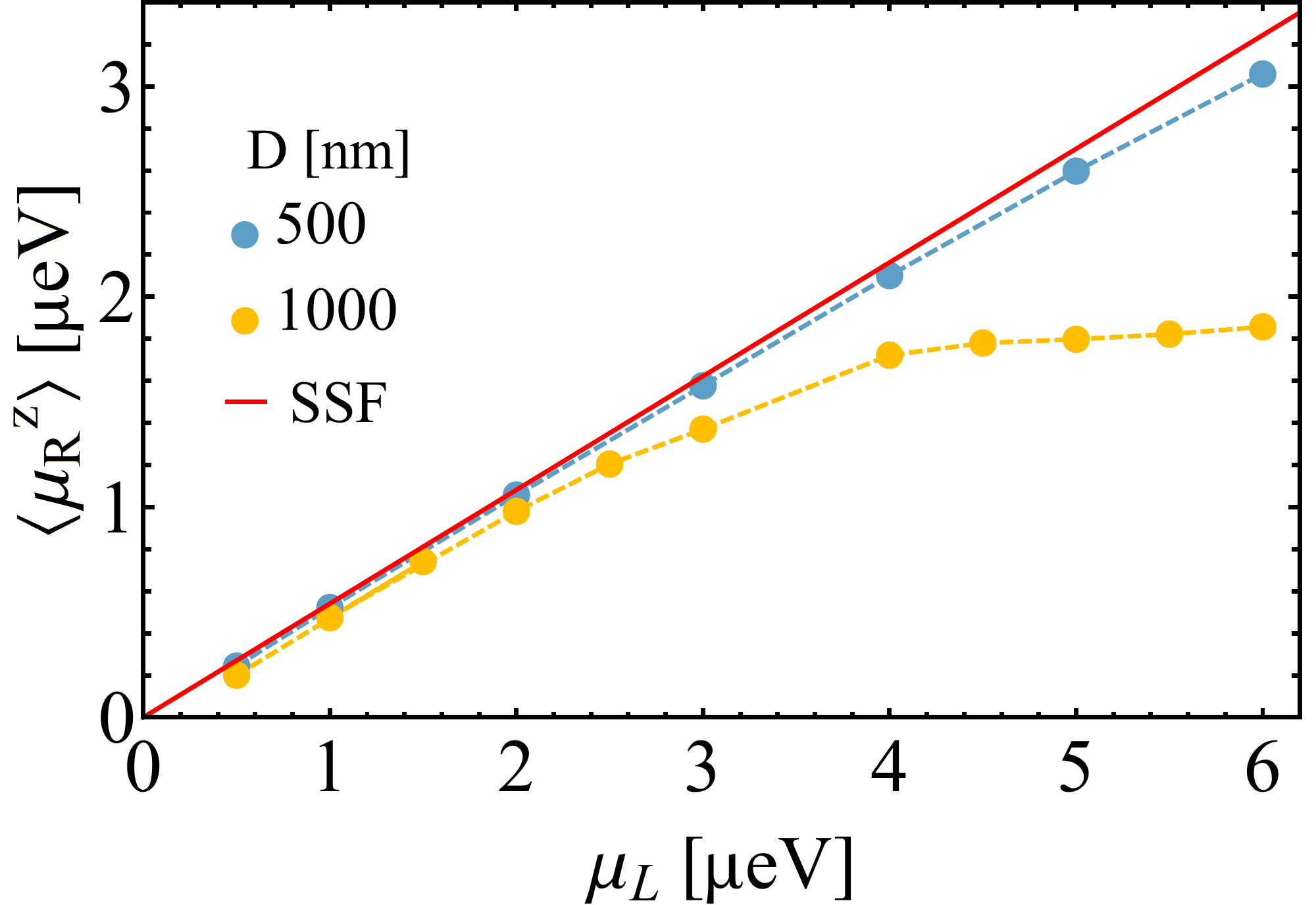}
\put(0,3){(c)}
\put(38,53){\includegraphics[width=1cm]{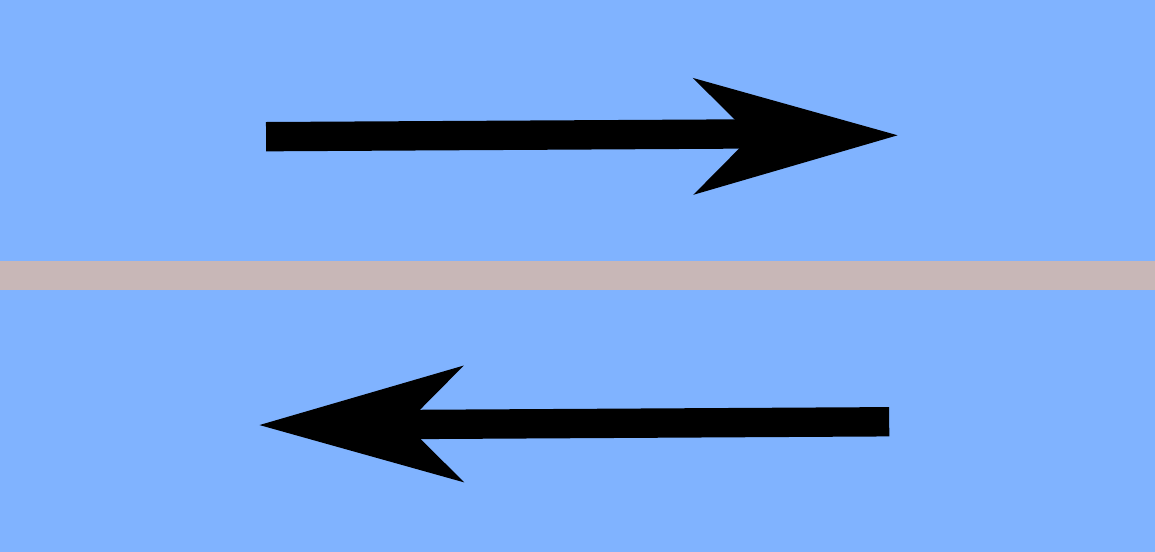}}
\end{overpic}
\begin{overpic}
[width=0.49\columnwidth,clip=true]{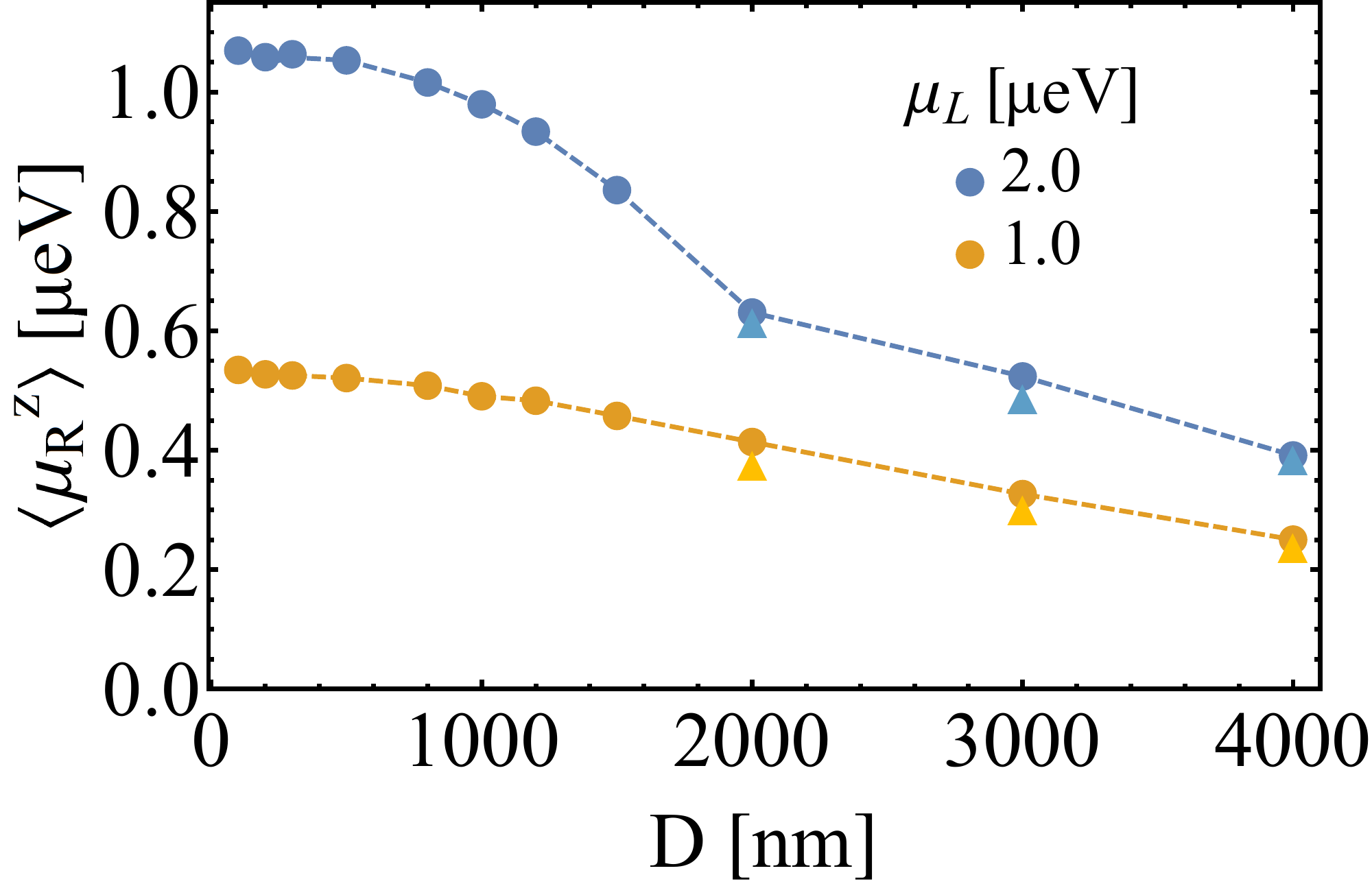}
\put(0,3){(d)}
\end{overpic}
\caption{(color online).
The average dc spin accumulation in the right contact, $\langle \mu^z_R \rangle$, as a function of (a) $\mu_L$ and (b) $D$ for a disk-shaped thin film.
$\langle \mu_R^z \rangle$ as function of (c) $\langle \mu^z_R \rangle$ and (d) $D$ for disk-shaped tri-layer samples.
The circles in panels (c) and (d) represent the results for symmetric tri-layer structures, where $M_{s,1/2}=M_s$, whereas the triangles in panel (d) indicate results for asymmetric layers with $M_{s,1/2}=(1\pm0.1)M_s$
The solid red lines in (a) and (c) represent the theoretical SSF values of $\langle \mu_R^z \rangle$.
}
\label{fig:plots disc}
\end{figure}

Next, we turn to FI thin films with circular geometries.
Such high-symmetry structures are chosen because
the absence of an easy axis leads to much longer spin-current propagation lengths.
The areas of the injection and detection contacts are 1/5th of the total area of the YIG disk (see Fig.~\ref{fig:system}(c)).
Fig. \ref{fig:plots disc}(a) shows that although spin transport is indeed possible for samples with diameters $D$ up to $1\,\mu$m, the SSF is still restricted to sample sizes $\lesssim 500$ nm and low values of $\mu_L$.
Fig. \ref{fig:plots disc}(b) similarly shows that $\langle \mu_R^z \rangle$ is independent of $D$ only for small values of $D$ and $\mu_L$.
In contrast with the square geometry, no dipole pinning occurs because of the lack of easy axes, and long-range non-SSF spin transport may occur in micrometer-sized samples.

Finally, we demonstrate that long-range SSF can be recovered in a synthetic antiferromagnet structure.
When two FI thin films are in contact via a thin normal metal, RKKY
interaction can lead to antiferromagnetic exchange coupling between the two layers \cite{parkin:prl991}. In the absence of an external magnetic field, the ground state has an antiparallel configuration with zero net magnetization. 
Dipolar interactions are suppressed over distances longer than the tri-layer thickness. Consequently, only the easy-plane anisotropy term survives 
in the thin-film limit.

By applying spin accumulation to the top FI in the same manner as for the single-layer FI, one can induce a rotation of the FIs' magnetizations that maintains the net magnetization close to zero. 
Under steady-state conditions, the SP current that flows from FI1 into the spacer layer is exactly compensated by the SP current that flows from FI2, thereby resulting in a vanishing 
SP+STT torque, and vice versa. Hence, the RKKY interaction dominates the interlayer interaction.
Writing the magnetizations as
$\mathbf{m}_{1/2}(\mathbf{r},t)=\left(\pm\sqrt{1-m_{1/2,z}^2}\cos\phi_{1/2},\pm\sqrt{1-m_{1/2,z}^2}\sin\phi_{1/2},m_{1/2,z}\right)$,
and assuming that $(m_{1,2})_z\ll 1$
and $\lvert \phi_1-\phi_2\rvert\ll1$, the SSF hydrodynamic equations for the first layer are
\begin{subequations}\label{eq:llg12}
\begin{eqnarray}
\dot{m}_{1,z}&=& \frac{2\gamma A}{M_{s,1}}\nabla^2 \phi_{1}-\alpha_{1} \dot{\phi}_{1} ,  \\ 
\dot{\phi}_{1}&=&[4 \pi M_{s,1} \gamma+\omega_\text{E,1}] m_{1,z}+\omega_\text{E,1} m_{2,z} \nonumber \\
&&+\alpha_{1} \dot{m}_{1,z}.
\end{eqnarray}
\end{subequations}
(For the second layer, interchange $1 \leftrightarrow 2$.)
Here, $M_{s,1(2)}$ is the saturation magnetization of layer 1(2) and 
$\alpha_{1}=\alpha_{0,1}+\alpha'_{1}$, where $\alpha_{0,1}$ is the intrinsic Gilbert damping in FI1 and $\alpha'_{1}=\alpha_{L(R)}$ under the left (right) contact area and zero otherwise. The second layer is not attached to any external contacts, i.e., $\alpha'_{2}=0$, and the damping is dominated by the layer's intrinsic Gilbert damping, $\alpha_2=\alpha_{0,2}$.
The strength of the RKKY interaction is given by
$\omega_\text{E,1(2)}=\gamma J/d_{2(1)}M_{s,2(1)}$, where $J>0$ is the interlayer
exchange-energy areal density \cite{gruenberg:prl86} and $d_{1(2)}$ is the thickness of layer 1(2).
The left and right contacts are attached to layer 1 and provide additional STT and SP, as in the single-layer cases.

For small values of $\boldsymbol{\mu}_L$, the spatial variation of $\mathbf{m}_{1,2}$ is small.
Assuming symmetric layers, i.e. $M_{s,1}=M_{s,2}=M_s$, $\alpha_{0,1}=\alpha_{0,2}=\alpha_{0}$, and $\omega_{\text{E},1}=\omega_{\text{E},2}=\omega_{\text{E}}$, we obtain 
\begin{equation}\label{eq:triOmega}
\langle \mu^z_R\rangle=-\hbar\Omega= \frac{g^\perp_L }{g^\perp_L+g^\perp_R+g_\alpha} \mu_L, 
\end{equation}
$m_z=-\Omega/(2\omega_\text{E}+4 \pi M_s \gamma)$, $(\phi_1-\phi_2)\sim \alpha_2 \Omega /\omega_\text{E}$, and $g_\alpha$ is the intrinsic conductance defined in Eq.~\eqref{eq:gintrinsic} with the replacement $\alpha_0 \rightarrow \alpha_{0,1}+\alpha_{0,2}$. Equation \eqref{eq:triOmega} can be identified as Ohm's law \eqref{eq:Ohm}.

Fig. \ref{fig:plots disc}(c) shows the exact numerical result for $\langle \mu_R^z \rangle$ as a function of $\mu_L$ for a tri-layer structure composed of two disks, each with thickness $d_{\text{YIG}}=5$ nm and coupling given by $\omega_E=7.3 \cdot 10^{10}$ s$^{-1}$, where the dipole interaction is fully included.
The detector signal is close to the ideal value given by Eq.~\eqref{eq:triOmega}, even for micrometer-sized systems.
Saturation also occurs in tri-layer systems but at much higher values of $\mu_L$ than for single-layer films; despite relatively large variations in $\phi$,
SSF remains stable because of screening of the dipole interactions.
Fig. \ref{fig:plots disc}(d) shows
that spin transport is possible over much greater distances in tri-layer structures than in single-layer ones:
the single-layer $\langle \mu_R^z \rangle$ exhibits a $75 \%$ spin-signal reduction over the $D$ interval $0.1 - 2.0 \, \mu$m whereas the tri-layer only experiences a $25 \%$ reduction over the same interval and a $50 \%$ reduction at $D=4\, \mu$m.
The SSF is robust against small variations in the FI layer properties; see Fig.~\ref{fig:plots disc}(d).
Our results demonstrate that tri-layer structures are able to support SSF currents for system sizes up to $\sim 1 \, \mu$m and long-range spin transport across samples several micrometers in size.

In summary, we have investigated SSF-mediated spin transport in FI thin films.
The dipole field qualitatively alters the transport properties so that single-layer SSF is possible only in systems less than a few hundred nanometers in size.
Suppression of the dipole field in tri-layer structures enables long-range spin transport mediated by SSF over length scales up to $\sim 1 \, \mu$m and by non-SSF magnetization dynamics over length scales up to several micrometers.

% % % % % % % % % % %
\acknowledgments
We gratefully acknowledge useful discussions with Peder Notto Galteland.
This work was supported by InSpin 612759.

%%%%%%%%%%%%%%%%%%%%%%%%%%%%%%%%%%%%%%%%%%%%%%%%%%%%%%%%%%%%%%%%%%%%%%%%%%%%%%%%%%%%%%%%%%%%%%%%%%%%%


\begin{thebibliography}{99}

\bibitem{konig:prl02} J. K\"onig, M. Chr. B\o nsager, and A. H. MacDonald, Phys. Rev. Lett. \textbf{87}, 187202 (2001).

\bibitem{nogueira:epl04} F. S. Nogueira and K.-H. Bennemann, Europhys. Lett. \textbf{67}, 620 (2004).

\bibitem{sonin:advphys10} E. B. Sonin,
Phys. Adv. Phys. \textbf{59}, 181 (2010). 

\bibitem{takei:prl14} S. Takei and Y. Tserkovnyak,
Phys. Rev. Lett. \textbf{112}, 227201 (2014).

\bibitem{chen:prb14} H. Chen, A. D. Kent, A. H. MacDonald, and I. Sodemann, Phys. Rev. B \textbf{90}, 220401 (2014).

\bibitem{takei:arxiv2015} S. Takei and Y. Tserkovnyak, arXiv:1506.01059.


\bibitem{nakata:prb14} K. Nakata, K. A. van Hoogdalem, P. Simon, and D. Loss, Phys. Rev. B \textbf{90}, 144419 (2014).

\bibitem{Clausen:arXiv1503.00482} P. Clausen, D. A. Bozhko, V. I. Vasyuscka, G. A. Melkov, B. Hillebrands, and A. A. Serga, arXiv:1503.00482.


\bibitem{hillebrands:apl14} P. Pirro, T. Br{\"a}cher, A. V. Chumak, B{\"a}gel, C. Dubs, O. Surzhenko, P. G{\"o}rnert, B. Leven and B. Hillebrands,
Appl. Phys. Lett. \textbf{104}, 012402 (2014).

\bibitem{kalinikos:jphys86} B. A. Kalinikos and A. N. Slavin, J. Phys. C \textbf{19}, 7013 (1986).

\bibitem{serga:jpd10} A. A. Serga, A. V. Chumak, and B. Hillebrands, J. Phys. D: Appl. Phys. \textbf{43}, 264002 (2010).

\bibitem{demokritov:nature06} S. O. Demokritov, V. E. Demidov, O. Dzyapko, G. A. Melkov, A. A. Serga, B. Hillebrands, and A. N. Slavin, Nature \textbf{443}, 430 (2006).

\bibitem{chumak:natcomm14} A. V. Chumak, A. A. Serga, and B. Hillebrands, Nature Comm. \textbf{5}, 1 (2014).


\bibitem{brataas:prl00} A. Brataas, Y. V. Nazarov, and G. E. W. Bauer, Phys. Rev. Lett. \textbf{84}, 2481 (2000).

\bibitem{gurevich:96} A. G. Gurevich and G. A. Melkov, Magnetic Oscillations and Waves, CRC, New York (1996).


\bibitem{vansteenkiste:14} A. Vansteenkiste, J. Leliaert, M. Dvornik, M. Helsen, F. Garcia-Sanchez, and B. Van Waeyenberge, AIP Advances \textbf{4}, 107133 (2014).

\bibitem{villamor:prb2013} E. Villamor, M. Isasa, L. E. Hueso, and F. Casanova, Phys. Rev. B \textbf{88}, 184411 (2013).

\bibitem{wang:prl14} H. L. Wang, C. H. Du, Y. Pu, R. Adur, P. C. Hammel, and F. Y. Yang, Phys. Rev. Lett. \textbf{112}, 197201 (2014).

\bibitem{du:prappl14} C. Du, H. Wang, F. Yang, and P. C. Hammel, Phys. Rev. Appl. \textbf{1}, 044004 (2014).

\bibitem{klingler:jpd15} S. Klingler, A. V. Chumak, T. Mewes, B. Khodadadi, C. Mewes, C. Dubs, O. Surzhenko, B. Hillebrands, and A. Conca, J. Phys. D: Appl. Phys. \textbf{48}, 015001 (2015).

\bibitem{suhl:pcs57} H. Suhl, Phys. Chem. Solids \textbf{1}, 209 (1957).

\bibitem{bahlmann:jap96} N. Bahlmann, R. Gerhardt, M. Wallenhorst, and H. D\"otsch, J. Appl. Phys. \textbf{80}, 3977 (1996).

\bibitem{parkin:prl991} S. S. P. Parkin, Phys. Rev. Lett. \textbf{67}, 3598 (1991).

\bibitem{gruenberg:prl86} P. Gr\"unberg, R. Schreiber, Y. Pang, M. B. Brodsky, and H. Sowers, Phys. Rev. Lett. \textbf{57}, 2442 (1986).




\end{thebibliography}
\end{document}